\documentclass{emulateapj}
\shorttitle{Fomalhaut b as a Dust Cloud}
\shortauthors{Lawler et al.}

\begin{document}

\title{Fomalhaut~b as a Dust Cloud: Frequent Collisions within the Fomalhaut Disk}

\author{S.~M.~Lawler\altaffilmark{1}\altaffilmark{2}, S.~Greenstreet\altaffilmark{3}, and B.~Gladman\altaffilmark{3}}
\altaffiltext{1}{Department of Physics and Astronomy, University of Victoria, PO Box 1700, STN CSC
Victoria, BC V8W 2Y2, Canada}
\altaffiltext{2}{NRC-Herzberg, 5071 West Saanich Road., Victoria, BC V9E 2E7, Canada}
\altaffiltext{3}{Department of Physics and Astronomy, University of British Columbia, 6224 Agricultural Road Vancouver, BC V6T 1Z1, Canada}

\begin{abstract}

The planet candidate Fomalhaut~b is bright in optical light but undetected in 
longer wavelengths, requiring a large, reflective dust cloud.
The most recent observations find an extremely eccentric
orbit ($e\sim0.8$), indicating that Fomalhaut~b cannot be the planet that is
constraining the system's eccentric debris ring.
An irregular satellite swarm around a super-Earth has been proposed, however,
explaining the well-constrained debris ring
requires an additional planet on an orbit that crosses that of the putative super-Earth.
This paper expands upon a second theory: Fomalhaut~b 
is a transient dust cloud produced by a catastrophic collision between 
planetesimals in the disk.
We perform collisional probability simulations of the Fomalhaut debris disk 
based on the structure of our Kuiper belt,
finding that the catastrophic disruption rate of $d\simeq100$~km bodies in the high-eccentricity
scattering component is several per decade.
This model paints a picture of the Fomalhaut system as having recently 
(within $\sim$10-100~Myr) experienced a dynamical instability within its planetary system,
which scattered a massive number of planetesimals onto large, high eccentricity orbits
similar to that of Fom~b.
If Fomalhaut~b is indeed a dust cloud produced by such a collision, we should soon see another appear,
while Fomalhaut~b will expand until it is either resolved or becomes 
too faint to be seen.

\end{abstract}

\section{Introduction}

Fomalhaut is a nearby (7.70~pc), widely-spaced triple star system \citep{Mamajeketal2013}
with an age of 440~Myr \citep{Mamajek2012}. 
Fomalhaut~A (Fomalhaut hereafter), is an A4 star that dominates the system 
and possesses a beautiful, eccentric debris ring that is resolved at optical \citep{Kalasetal2005}, 
infrared \citep{FajardoAcosta1998,Stapelfeldtetal2004,Suetal2013,Ackeetal2012}, 
submillimeter \citep{Hollandetal1998,Hollandetal2003,Marshetal2005}, and 
millimeter wavelengths \citep{Boleyetal2012}.

Fomalhaut~b was directly imaged using the {\it Hubble Space Telescope} 
\citep{Kalasetal2008} and, based on two observation epochs, 
plausibly consistent with orbital predictions for a massive planet dynamically constraining the 
eccentric dust ring \citep{Quillen2006,Chiangetal2009}.
However, follow-up observations in infrared wavelengths failed to detect Fom~b
\citep{Marengoetal2009,Jansonetal2012}, indicating a significantly lower mass than Jupiter.

The nature of Fom~b is uncertain.
The most recent observations of Fom~b suggest that it is on a highly eccentric,
possibly ring-crossing orbit \citep{Kalasetal2013,Beustetal2014}.
Fom~b is also brighter at optical wavelengths than predicted by planetary atmosphere models,
requiring a cloud of optically reflective small dust grains.  
\citet{KennedyWyatt2011} suggest a swarm of collisionally 
grinding irregular satellites, gravitationally bound to a super-Earth planet.
However, dynamical analyses suggest that a planet of this mass on such an eccentric orbit will destroy
the apsidally aligned debris ring within a very small fraction ($<$1\%) of the stellar age
\citep{Kalasetal2013,Tamayo2014,Beustetal2014}.

Several papers have discussed an alternate option, that Fom~b is a post-collision
dust cloud \citep{Currieetal2012b,Galicheretal2013,Kalasetal2013,Tamayo2014,Kenyonetal2014},
and each dismissed this theory because the collisional timescales between planetesimals
appears to be uncomfortably long.  
However, examination of our own Kuiper Belt's structure 
and an understanding of the clearing of our own Solar System
suggests that a significant, unaccounted for, high-eccentricity component may be present.

In this paper, we show that Fom~b is plausibly a collision-generated dust cloud, 
finding a catastrophic disruption rate for objects on Fom~b-like orbits high enough that
observing a post-collision dust cloud is tenable.

\section{Comparison with the Kuiper Belt} \label{sec:kuiper}

\begin{figure}
\centering
\includegraphics[scale=0.35]{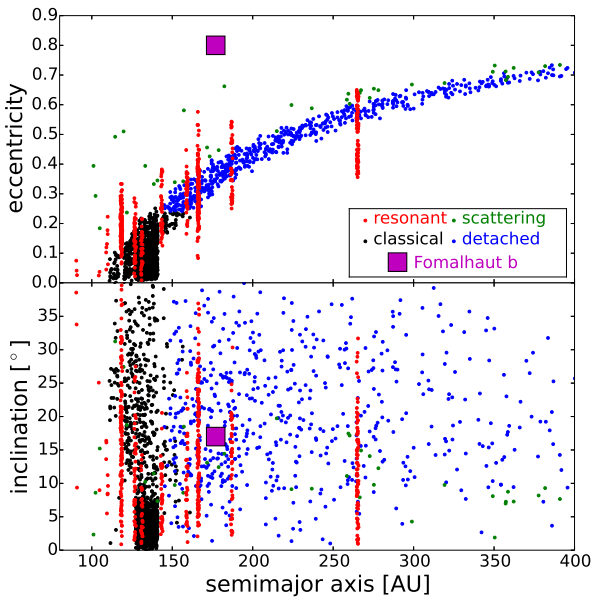}
\caption{
Orbital element distributions for the debiased Kuiper Belt as measured by CFEPS,
where the semimajor axis has been scaled so that the classical KBOs match
Fomalhaut's main debris ring.
The different populations are shown by different colors, labelled in the legend. 
Also shown on the plot is the best-fit orbital elements for Fom~b 
\citep[][large purple square]{Kalasetal2013}.
Fomalhaut~b most clearly belongs in the scattering population.
}
\label{fig:cfeps}
\end{figure}

As \citet{Kalasetal2013} point out, Fomalhaut's debris ring
and our Solar System's main Kuiper belt are similar dynamically, despite their different ages.
A perhaps important difference is that while the classical Kuiper belt has very low eccentricity,
the Fomalhaut debris ring has a significant forced eccentricity of $e\simeq0.1$.
Looking at the large (diameter $d\gtrsim100$~km) objects in the 
Kuiper belt, the only apparent structure
is the dense, relatively narrow ring of the main classical belt \citep{LawlerIAU}.
Due to the high optical depth of the dust grains in the Fomalhaut ring 
($\sim$10$^{-3}$; \citealt{Marshetal2005}) compared to the Kuiper Belt ($\sim$10$^{-7}$),
the maximal dust distribution should co-located with the ring 
(see simulations by \citealt{KuchnerStark2010}).

Fom~b's orbit is placed in context in Figure~\ref{fig:cfeps},
where the Kuiper belt, as measured by the Canada-France Ecliptic Plane Survey
\citep[CFEPS;][]{PetitL7,GladmanReson}
had its semimajor axis multiplied by three to match the Fomalhaut system.
Though many KBO dynamical sub-classes exist, here (Figure~\ref{fig:cfeps}) 
we only discuss four broad categories: 
resonant objects (red) which are in mean-motion resonances with Neptune,
scattering objects (green) whose orbits significantly change due to 
close encounters with the giant planets within a few Myr, 
detached objects (blue) which have eccentricities correlated with their semimajor axes, implying
past planetary interactions, and lastly
classical objects (black) which include the rest of the KBOs
\citep[see][]{Gladmanetal2008}.
The resonant semimajor axes in the Kuiper Belt depend
on the location of Neptune, thus we do not expect to find objects in these exact locations in
the Fomalhaut disk.

With a pericenter well inside the ring,
Fom~b is clearly a member of the scattering population.
Our hypothesis is that Fom~b is a newly disrupted member of Fomalhaut's 
scattering planetesimal population, in a recently perturbed Fomalhaut system,
similar to the Nice Model instability that occurred early in our Solar System's history.

\section{Modelling the Fomalhaut Disk} \label{sec:scaledcfeps}

Our model of the Fomalhaut disk is based on the known structure and 
populations of the Kuiper Belt.
The Kuiper Belt's scattering disk in particular is heavily eroded from its state just after the
migration of the giant planets.
Simulations show that today's scattering population is $\sim$1\% of the original
\citep{DuncanLevison1997}, which has been eroded away mainly
by planetary close encounters causing ejection from the Solar System.
Because Fomalhaut is significantly younger than the Sun,
in order to transform our Kuiper Belt into an approximation of the Fomalhaut disk, 
we first must correct for this erosion.
This brings the number of scattering
objects to about the same as the classical population, though in a much more widely
dispersed orbital distribution.
Additionally, very massive planets ($\gtrsim5$~$M_{\rm Jup}$) with projected separations $>15$~AU
have been ruled out \citep{Currieetal2013},
so the timescale for depletion of the scattering population shouldn't
be much faster than in our Solar System.
Thus, the number of scattering objects is scaled up by 100 times relative to other
Kuiper Belt populations.

The orbital distribution of objects in the 
young scattering disk was different 
from today's distribution, as some orbits are more likely to be ejected than others.
We produced a simple ``young scattering'' distribution for our Solar System 
by starting with a flat distribution of objects
with $a=4-35$~AU, $e=0.1$, and $i=15^{\circ}$, and running a numerical integration with the four giant
planets for 100~Myr.  
The surviving objects, many of which show more deeply crossing perihelia than for the
scattering population shown in Figure~\ref{fig:cfeps},
had their semimajor axes tripled to match 
the Fomalhaut system.

Next, one must also scale up the system mass until the main classical belt matches the Fomalhaut
ring's much larger observed mass compared to our Kuiper Belt.
CFEPS measured the absolute number of $d\gtrsim100$~km Kuiper Belt objects, so
matching the mass of another disk is accomplished by scaling the number of 
$d>100$~km objects in the model.
Coincidentally, $d\simeq100$~km is the minimum size required to produce the observed Fom~b dust cloud
\citep{Galicheretal2013,Kenyonetal2014}.
We settled on a scaling factor of 1000. 
This is very approximate,
as mass estimates for the Kuiper Belt vary by an order of magnitude 
\citep[$\sim$0.01-0.1~$M_\oplus$; e.g.,][]{Fraseretal2014,Vitenseetal2010}, and mass estimates for the
Fomalhaut ring vary by two orders of magnitude 
\citep[$\sim$2-110~$M_\oplus$; e.g.,][]{Boleyetal2012,Ackeetal2012}. 

Orbital elements for simulated planetesimals in the Fomalhaut model are
based on the CFEPS L7 synthetic model of the Kuiper Belt (available at {\tt www.cfeps.net}).
The heart of the classical Kuiper Belt sits at $\sim$45~AU, 
while the peak of the Fomalhaut disk as measured by ALMA \citep{Boleyetal2012} is at $\sim$135~AU,
so we multiplied the semimajor axes of each synthetic object by three.
To match the main ring shape and higher eccentricity, the eccentricities of the main classical belt 
were increased by 0.05,
so that the median eccentricity became $e\simeq0.1$.
To match the apsidal aligment of the planetesimals relative to each other, 
the simulated classical objects were set with $\Omega=-\omega\pm10^{\circ}$.

This orbital distribution of synthetic objects 
is the starting point for our collisional probability simulations.

\subsection{Collisional Probability Simulations of the Fomalhaut System}

We use an Opik collision probability code based on \citet{Donesetal1999}
to compute the 
collision probability for a Fom~b progenitor 
(radius $r=50$~km, $a=177$~AU, $e=0.8$, $i=17^{\circ}$) with various small body populations,
ignoring any other planets in the system. 
The code calculates the collision probability 
for a population of projectiles by numerically integrating over the 
precession cycle of the nodal longitude $\Omega$ and argument of pericenter $\omega$
for both the impactor and target bodies. 

The code was modified to bin the collision probability into individual impact 
velocity bins (as opposed to binning the total collision probability into an 
average impact velocity bin computed from all possible impact orientations over 
a full precession cycle of the orbital nodes),
as well as individual impact distance bins. 
This produces detailed impact 
velocity distributions for each projectile population.

\subsection{Catastrophic Collisional Probability}

Using the probabilities for a range of impact velocities in each 
distance bin, we calculate the total 
catastrophic disruption probability. 
Closer to the star, relative speeds are generally higher allowing 
smaller objects to produce the same collision energy as a slower, larger object 
farther from the star.

Equation~2 from \citet{LeinhardtStewart2012} relates target mass,
projectile mass, collisional velocity, and the catastrophic disruption energy for an icy 
body \citep[which we take from Figure~11 in][using $r=50$~km]{LeinhardtStewart2009}.
This gives the minimum projectile size
required to catastrophically disrupt a $d=100$~km target body at a given velocity.
To properly weight the collisional probabilities in each velocity bin, 
we multiply by the number of bodies of this minimum size, 
assuming a projectile size distribution of $dn/dr \propto r^{-q}$, with $q=3.5$,
as for a collisional cascade \citep{Dohnanyi1969},
consistent with other estimates of the disk mass \citep[e.g.,][]{Boleyetal2012,Ackeetal2012}
from the literature.

\begin{figure}
\centering
\includegraphics[scale=0.45]{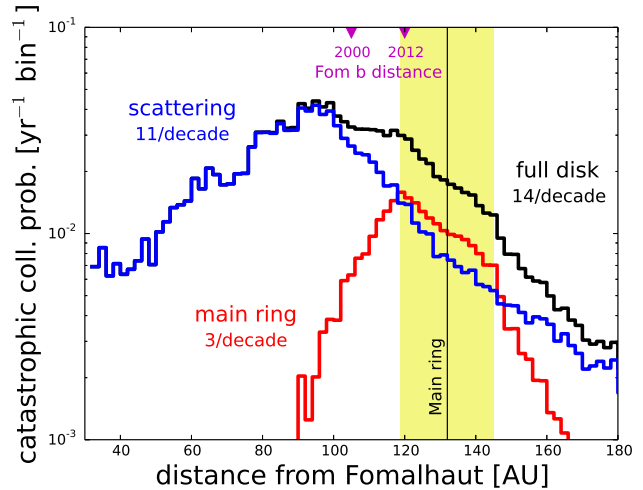}
\caption{
The catastrophic collision probability per year per 2~AU-wide bin for different projectile populations
onto a target with Fom~b's current orbit.
Black shows the full disk model, red shows just the non-scattering component (the main ring),
and blue shows just the scattering component, with total integrated catastrophic disruption
probabilities stated for each population. 
Yellow band shows the pericenter-apocenter range of main ring particles, the mutual main ring
collision rate (not shown) is a factor of several larger.
Measured stellocentric distances of Fom~b in 2000 and 2012 are shown.
}
\label{fig:probvsdistance}
\end{figure}

Figure~\ref{fig:probvsdistance} shows the catastrophic collision probability distribution
per year per 2~AU-wide bin, for three different projectile populations.
Integrating over each curve gives the total probability per year
of a given population catastrophically disrupting a $d=100$~km progenitor on a Fom~b-like orbit.

The full disk collisional model (black line in Figure~\ref{fig:probvsdistance}) 
shows the likelihood of collisions
between an object on a Fom~b-like orbit and anything in the Fomalhaut system model.
The total integrated catastrophic collisional probability is 14/decade, and
the most likely place for collisions is just sunward of the main ring. 
The portion of the catastrophic probability for 
just the non-scattering objects is also shown separately (red line);
as expected the main ring dominates at its distance,
but most scattering object disruptions are mutual and occur between 50-110~AU.
We also confirmed that our model produces a catastrophic disruption rate for collisions
{\it between objects in the main ring} that approximately 
reproduces the observed dust production rate \citep{Ackeetal2012}.

Though the probabilities for disruption of a Fom~b-like orbit are high within the main belt, 
collisions here would not generate a dust cloud
because the high dust density within the ring would effectively absorb and/or
collisionally destroy the dust cloud almost instantly.
Also, due to Fom~b's orbital direction, we know Fom~b has not been inside the main belt for 
at least $\sim$250~years.
Therefore, this is not our favoured projectile population.

The most likely population to cause a catastrophic collision is the scattering population
(blue line in Figure~\ref{fig:probvsdistance}).
This curve peaks at about 95~AU from Fomalhaut, similar to Fom~b's distance $\sim$10-15 years ago,
with a total integrated catastrophic collision rate for $d=100$~km objects of 11/decade.
The most likely collision is a $\sim$10~km projectile destroying a 100~km target.

\section{Plausibility of Fom~b as a Dust Cloud}

\citet{Galicheretal2013} estimate that within the main belt,
mutual collisions between $d=100$~km bodies happen once per century 
(they do not calculate catastrophic disruption rate),
while \citet{Currieetal2012b} and \citet{Kalasetal2013} 
argue that the dust cloud scenario is unlikely because of the 
long collisional timescale and relatively quick orbital shearing.
Our analysis shows that closer to the star than the main
ring, higher mutual collision speeds result in smaller objects being able
to catastrophically disrupt the same size target.  
Because of the size distribution, there are abundantly more smaller objects,
which shifts the peak destruction probability sunward of the main ring.

\citet{Kenyonetal2014} use size distribution models to argue that a dust cloud resulting from
a collision is only possible for a limited range of parameters.
A 50~km radius progenitor can reproduce the observed properties for 
a dust power-law size index ($q$) of 4.7 or less, with the largest remaining fragment less than
1\% the size of the initial body (otherwise there is not enough mass in dust to reproduce observations).
While similar $q$ values are often reproduced in simulations of catastrophic disruptions,
near-complete destruction of the target is only rarely produced in simulations \citep{LeinhardtStewart2012}.
Additionally, \citet{Kenyonetal2014} point out that the expansion of the dust cloud is
constrained by the eight year observational baseline.
If the dust cloud is unresolved \citep{Currieetal2012b,Kalasetal2013,Kenyonetal2014},
this limits the expansion to $\lesssim$300~m/s, while if it is marginally resolved
\citep{Galicheretal2013}, the expansion is limited to $\lesssim$1~km/s.
Both of these values are significantly less than the relative velocities for the
bulk of simulated collisions.

However, as discussed in \citet{Kenyonetal2014},
the expansion rate of a dust cloud resulting from a catastrophic collision
onto a fairly large body ($\sim$100~km) depends not on the collision velocity,
but on the escape velocity of the total mass \citep[projectile+target;][]{BenzAsphaug1999}.
For a $d=100$~km icy body, the escape velocity is approximately 30~m/s, suggesting
that there is close to a century before the dust cloud would expand enough to be resolved
by {\it Hubble}, greatly increasing the window of time the dust cloud remains unresolved.

This theory implies that there are many possibly visible dust clouds in the Fomalhaut system
at any given time.  
However, we note the difficulty of these observations: only a limited range of distances 
and longitudes have deep enough magnitude limits to detect such a cloud 
\citep{Kalasetal2013,Currieetal2012b}.
Also, only a fraction of the catastrophic
collisions will result in a visible dust cloud \citep{Kenyonetal2014}.
Lastly, 11/decade is probably an overestimate since our scattering
population estimate is appropriate for just after the system's 
perturbation; if instead the perturbation occurred $\sim$30 Myr ago 
(perhaps a more reasonable $\sim$10\% of the system's age) our integrations show that 
the catastrophic collision rate for the scattering population would
drop to $\sim$1/decade as the unseen planets gradually eject planetesimals.

A recent example of an observed expanding cloud involves
Comet 17P/Holmes, which outburst in 2007 resulting in an
optically bright, nearly spherical cloud of dust.
While this outburst was most likely due to a gas-pressure driven explosion and not
a collision \citep[e.g.,][]{Stevensonetal2010}, it provides a well-studied analogue to
the sort of expanding dust cloud we propose to explain Fom~b.
The cloud of Comet Holmes was roughly divided into two parts with similar masses 
that expanded at different rates \citep{Reachetal2010}:
the smallest ($\sim$2~$\mu$m) grains expanded the fastest ($\sim$250~m/s), 
creating a large shell that was the most easily visible, 
and a second cloud composed
of larger ($\sim$200~$\mu$m) grains expanding more slowly ($\sim$9~m/s).
Because of Fomalhaut's higher luminosity, the blowout limit ($\sim$10~$\mu$m for icy grains)
is much larger than in our Solar System.
Any grains smaller than this would immediately be placed on hyperbolic orbits.
The relative fraction of large versus small dust grains produced by the collision
would dictate which portion would be more easily visible: either the larger grains
that continue on roughly the same orbit as the parent body, or the small grains which 
find themselves on either an extremely elliptical or even hyperbolic orbit.

The latter offers another possible explanation for the extreme orbit of Fom~b,
as small grains with high $\beta$ values will instantaneously have their
orbits shift to high $e$.  
While \citet{Beustetal2014} rule out $e>0.98$ in their orbital fits for Fom~b,
\citet{Kalasetal2013} do not specifically rule out unbound orbits.
If Fom~b's orbit is found to be unbound in future observations, 
a dust cloud made of small grains is a more plausible explanation
than an extremely fortuitous observation of a planet being ejected from the system.

\section{Discussion: Evolution of the Fomalhaut System}

The theory that Fom~b is a collisional dust cloud fits into the following possible narrative
of the Fomalhaut system.
Fomalhaut hosts a young planetary system.
As has been proposed for the early few hundred Myr of our own Solar System \citep{Gomesetal2005}, 
the Fomalhaut system experienced a relatively recent ($\sim$100~Myr)
dynamical instability that resulted in a dramatic reshuffling of the orbits of its planets
and smaller bodies.
Simulations of our own Solar System show that a dynamically cold population (like the debris ring)
can survive such a violent event \citep{Batyginetal2011}.
Fomalhaut~c (as yet unobserved) was scattered or migrated 
onto its current moderately eccentric orbit ($e\simeq0.1$)
and is currently secularly driving the eccentricity of the main debris ring, as
discussed in previous works.
Any planetesimals that were on orbits interior to the main ring ($a\simeq135$) were scattered, 
and this scattering population is currently being eroded by close encounters with Fom~c
and any other planets interior to Fom~c.
The erosion rate of this population depends on the masses and orbital configurations of
the planetary system: 
larger planets will erode the scattering disk faster,
as will more closely spaced planets, since they are more easily able to ``hand off'' scattered
planetesimals to each other.

Due to the low surface density of this scattering population relative to the main ring,
a belt of dust resulting from mutual collisions within this population would not be as easily visible,
and indeed much of it could be ejected from the system immediately after formation
due to high $\beta$ values \citep[e.g.,][]{Wyatt2008}.
Thus, Fomalhaut could be hiding a significant population of KBO-analogues
on highly eccentric orbits similar to that of Fomalhaut~b.
More detailed modelling is required to determine the orbital population and 
spectral energy distribution of dust resulting from these collisions within the scattering population.
A likely outcome would be a population of warm dust grains, as found by
recent IR observations \citep{Suetal2013,Ackeetal2012}.

There are two predictions of our Fom~b-as-a-dust-cloud hypothesis.
The first is that the $\sim$30-100~m/s Fom~b expansion 
over the coming decades will eventually permit the object to be resolved.
The second prediction is that more Fom~b-like dust clouds should be created in the system 
in the near future. 

\section{Summary and Conclusion}

We have shown that the catastrophic disruption rates within the Fomalhaut debris disk can be much higher
than previously assumed. 
This is primarily due to inclusion of a high eccentricity scattering
component, equivalent to the scattering population in our 
Solar System's Kuiper Belt. 
Fomalhaut's young age is also an important factor
permitting a more massive scattering component.
This calculation shows that the possibility that Fom~b is just a cloud of dust should not be dismissed.

The relatively high collision rate that we calculate here would
mean that another Fom~b-like object should appear within the next decade,
and Fom~b itself will fade over the coming years, possibly becoming resolved.
In order to test these two predictions, continued follow-up observations
capable of detecting objects as faint as Fom~b are vital.
For now, the only telescope capable of detecting Fom~b is {\it Hubble},
but the upcoming {\it James Webb Space Telescope} will be able to resolve the dust cloud,
and provide some additional constraints on the dust composition with 
near-IR measurements.

While perhaps disappointing to think of Fom~b as merely a cloud of dust and not an actual planet, 
this scenario tells us about the 
structure of the Fomalhaut debris disk, and strongly implies the presence of multiple planets,
just as the scattering component in our Solar System is constrained and limited by the presence
of giant planets interior to the Kuiper Belt.

\acknowledgments {
S.M.L. wishes to thank Wes Fraser, Kat Volk, Aaron Boley, Yanqin Wu, Dan Tamayo,
Christa Van Laerhoven, and Henry Ngo
for enlightening discussions and advice that made this project possible.
S.M.L. acknowledges an NSERC Discovery Accelerator Supplement which funded this work.
}


\end{document}